\documentclass[preprint,12pt]{elsarticle}

\usepackage{amssymb}
\usepackage{lineno}
\usepackage{graphicx}
\usepackage{xcolor}
\usepackage{float}
\usepackage[autostyle=false, style=english]{csquotes}
\MakeOuterQuote{"}

\journal{Radiation Measurements}

\begin{document}

\begin{frontmatter}



\title{Study on the reusability of fluorescent nuclear track detectors using optical bleaching}


\author[inst1,inst2]{Abdul Muneem}

\affiliation[inst1]{organization={High Energy Nuclear Physics Laboratory, Cluster for Pioneering Research, RIKEN},
            addressline={2-1 Hirosawa}, 
            city={Wako},
            postcode={351-0198}, 
            state={Saitama},
            country={Japan}}

\affiliation[inst2]{organization={GIK Institute of Engineering Sciences and Technology},
            addressline={Topi}, 
            city={Swabi},
            postcode={23640}, 
            state={KP},
            country={Pakistan}}
\author[inst1,inst3]{Junya Yoshida}

\affiliation[inst3]{organization={Department of Physics, Tohoku University},
            addressline={Aramaki}, 
            city={Aoba-ku},
            postcode={980-8578}, 
            state={Sendai},
            country={Japan}}
\author[inst1]{Hiroyuki Ekawa}
\author[inst4]{Masahiro Hino}
\affiliation[inst4]{organization={Institute for Integrated Radiation and Nuclear Science, Kyoto University},
            addressline={Kumatroi}, 
            city={Sennan-gun},
            postcode={590-0494}, 
            state={Osaka},
            country={Japan}}
\author[inst5]{Katsuya Hirota}
\affiliation[inst5]{organization={High Energy Accelerator Research Organization (KEK)},
            addressline={Tsukuba}, 
            postcode={305-0801}, 
            state={Ibaraki},
            country={Japan}}    
\author[inst5,inst14]{Go Ichikawa}      
\affiliation[inst14]{organization={Japan Proton Accelerator Research Complex (J-PARC) Center},
            addressline={Tokai},
            postcode={319-1195}, 
            state={Ibaraki},
            country={Japan}}
\author[inst1,inst6]{Ayumi Kasagi}      
\affiliation[inst6]{organization={Graduate School of Engineering, Gifu University},
            postcode={501-1193}, 
            state={Gifu},
            country={Japan}}
\author[inst7,inst8]{Masaaki Kitaguchi}      
\affiliation[inst7]{organization={Department of Physics, Nagoya University},
            addressline={Furo-cho}, 
            city={Chikusa},
            postcode={464-8602}, 
            state={Nagoya},
            country={Japan}}
\affiliation[inst8]{organization={Kobayashi-Maskawa Institute for Origin of Particles and the Universe(KMI), Nagoya University},
            addressline={Furo-cho}, 
            city={Chikusa},
            postcode={464-8602}, 
            state={Nagoya},
            country={Japan}}
            
\author[inst17]{Satoshi Kodaira}
\affiliation[inst17]{organization={Space Quantum Research Group, QST Advanced Study Laboratory, National Institutes for Quantum and Radiological Science and Technology},
            postcode={263-8555}, 
            state={ Chiba},
            country={Japan}}

\author[inst5,inst14]{Kenji Mishima}

\author[inst2,inst9]{Jameel-Un Nabi}
\affiliation[inst9]{organization={University of Wah},
            addressline={Quaid Avenue}, 
            city={Wah Cantt},
            postcode={47040}, 
            state={Punjab},
            country={Pakistan}}

\author[inst1]{Manami Nakagawa}

\author[inst10]{Michio Sakashita}

\affiliation[inst10]{organization={RIKEN Center for Advanced Photonics, RIKEN},
            addressline={2-1 Hirosawa}, 
            city={Wako},
            postcode={351-0198}, 
            state={Saitama},
            country={Japan}}
\author[inst10]{Norihito Saito}            
            
\author[inst1,inst11,inst12]{Takehiko R. Saito}          
\affiliation[inst11]{organization={GSI Helmholtz Centre for Heavy Ion Research},
            city={Planckstrasse 1},
            postcode={64291}, 
            state={Darmstadt},
            country={Germany}}
\affiliation[inst12]{organization={School of Nuclear Science and Technology, Lanzhou University},
            addressline={222 South Tianshui Road}, 
            city={Lanzhou},
            postcode={730000}, 
            state={Gansu Province},
            country={China}}
            
\author[inst10]{Satoshi Wada}            
\author[inst13]{Nakahiro Yasuda}
\affiliation[inst13]{organization={Department of Nuclear Power Disaster Prevention and Risk Management, Research Institute of Nuclear Engineering, Fukui University},
            addressline={Tsuruga-shi}, 
            postcode={730000}, 
            state={Fukui-ken},
            country={Japan}}

\begin{abstract}
Fluorescent nuclear track detectors (FNTDs) based on Al${_2}$O${_3}$:C,Mg crystals are luminescent detectors that can be used for dosimetry and
detection of charged particles and neutrons. These detectors can be utilised for imaging applications where a reasonably high track density, approximately of the order of 1 $\times$ $10^4$ tracks in an area of 100 $\times$ 100 $\mu$m$^2$, is required. To investigate the reusability of FNTDs for imaging applications, we present an approach to perform optical bleaching under the required track density conditions. The reusability was assessed through seven irradiation-bleaching cycles. For the irradiation, the studied FNTD was exposed to alpha-particles from an $^{241}$Am radioactive source. The optical bleaching was performed by means of ultraviolet  laser light with a wavelength of 355 nm. Three dedicated regions on a single FNTD with different accumulated track densities and bleaching conditions were investigated. After every irradiation-bleaching cycle, signal-to-noise ratio was calculated to evaluate FNTD performance. It is concluded that FNTDs can be reused at least seven times for applications where accumulation of a high track density is required.

\end{abstract}




\end{frontmatter}


\section{Introduction}
\label{sec:sample1}
Fluorescent nuclear track detectors
(FNTDs) are luminescent solid-state detectors based on single crystals of aluminum oxide (Al${_2}$O${_3}$) doped with carbon and magnesium (Al${_2}$O${_3}$:C,Mg )\cite{Akselrod20006a, AKSELROD201835}. FNTDs have been successfully applied for neutron dosimetry \cite{AKSELROD2006295,SYKORA2008a1017,Akselrod2013}, as criticality
dosimeters \cite{Harrison2017}, for absorbed dose measurements on photon fields \cite{Akselrod20006a,Harrison2017}, in micro-beam radiation therapy \cite{BrauerKrisch2010},for single-track
measurements of protons and heavier ions \cite{ SYKORA2008a,Sawakuchi2016,BARTZ201424}, and in radiobiology
experiments \cite{NIKLAS20131141,Niklas2013a,Kodaira2015,Niklas_2016,OT10996,Kouwenberg2018}. FNTDs are small sized and chemically stable. In addition, FNTDs
allow to image objects with a micrometer spatial resolution. For these reasons FNTDs may also be applied for microdosimetry in boron neutron capture therapy and neutron imaging applications.

FNTDs have the capability to record three-dimensional trajectories of charged particles as they travel through these detectors. Incident charged particles induce ionization events within the crystal and produce electron-hole pairs in the valance band. The freed electrons move through the conduction band until
being trapped by the color centers (aggregate oxygen vacancy defects present in the crystal)
nearby the trajectory of the charged particles. This process results in photochromic
transformation of the color centers \cite{AKSELROD2006295,SYKORA2010631}. Consequently, the tracks of charged particles are recorded in the crystal. These tracks can be imaged using high resolution non-destructive readout systems based on confocal
laser scanning microscopy \cite{AKSELROD2006295, Akselrod2014}. The recorded tracks (also known as fluorescent tracks) are analogous to the tracks recorded in other tracking detectors, e.g. CR-39 plastic nuclear track detectors \cite{KHAN1983129} and nuclear emulsion \cite{Ariga2020}.

The radiation-induced fluorescent tracks recorded in FNTDs can be erased by employing heat treatment (thermal annealing) or using ultraviolet (UV) laser light (optical bleaching). The fluorescent tracks are stable up to 600 $^\circ$C, but can be erased by heating the crystals above 650 $^\circ$C \cite{AKSELROD201835, Akselrod2003}. Sykora \emph{et al.}, successfully erased FNTDs by employing thermal annealing at 680 $^\circ$C. Reusability, up to nine times, has been confirmed
by irradiating detectors to alpha-particles and thermal annealing \cite{SYKORA2008a1017}. Compared to thermal treatment, optical bleaching can reduce the background fluorescence more effectively and makes the background fluorescence more uniform, which results in a higher signal-to-noise ratio (SNR), where the SNR is defined as the ratio of the mean value of the fluorescence intensity of tracks created by alpha-particles in the single crystals to the standard deviation of the background \cite{AKSELROD201835, SYKORA2010631}. FNTDs can be optically bleached using UV lasers with wavelengths of 349 nm \cite{AKSELROD201835} and 325 nm \cite{SYKORA2010631}, through a two-photon absorption photoionization process. Sykora \emph{et al.}, used a 325 nm laser system consisting of an optical parametric oscillator pumped by tripled Nd:YAG laser for the photochromic transformation of the aggregate defects present in FNTDs. It was focused
in a spot of 0.12 mm$^2$, and uniform bleaching was achieved by using laser light raster
scanning on the FNTD \cite{SYKORA2010631}.

Along with reusability, FNTDs have several other merits as follows. FNTDs do not require post-irradiation chemical processing like other tracking detectors such as CR-39 plastic nuclear track detectors \cite{KHAN1983129} and nuclear emulsion \cite{Ariga2020}. FNTDs can detect charged particles in a wide
Linear Energy Transfer range and possess superior spatial resolution over CR-39 plastic nuclear track detectors \cite{AKSELROD2006295}.
FNTDs exhibit high sensitivity and high detection efficiency for heavy charged particles and dosimetry of neutrons as compared to that already existing nuclear track detectors \cite{AKSELROD201835, AKSELROD2006295, AKSELROD2012, MAKSELROD2006, AKSELROD20111671}. We plan to reap from these properties
and utilize FNTDs for micrometer scale neutron imaging applications.

FNTDs can be combined with neutron converters based on boron, lithium or gadolinium for neutron imaging applications. Hirota  \emph{et al.} performed neutron imaging of a gold bonding wire with a diameter of 30 $\mu$m in the crystal oscillator chip using fine-grained nuclear emulsion detector combined with boron based neutron converter under the accumulated track density of approximately 1 $\times$ $10^4$ tracks per 100 $\times$ 100 $\mu$m$^2$ \cite{jimaging7010004}. 

Since the reusability is one of the key advantages of FNTDs, in this paper we investigate the reusability of FNTDs under the accumulated track density of approximately 1 $\times$ $10^4$ tracks per 100 $\times$ 100 $\mu$m$^2$. The reusability was assessed through seven cycles of irradiation with alpha-particles and optical bleaching.
To accumulate the required track density for the imaging applications, a single FNTD was irradiated with alpha-particles from an $^{241}$Am radioactive source. After the irradiation, the regions with high accumulated track density were optically bleached using UV laser light at a wavelength of 355 nm. The irradiation with alpha-particles and optical bleaching were repeated seven times and the SNR values were calculated after each cycle to quantify the image quality of fluorescence images.

\section{Materials and methods}
\subsection{Overview}
The FNTD (8.0 $\times$ 4.0 $\times$ 0.5 mm$^3$ single crystal) used in this study was produced by
Landauer Inc., (Landauer Crystal Growth Division, Stillwater, OK, USA).
One large surface of the FNTD was polished. Prior to the optical bleaching, the FNTD was irradiated with alpha-particles from an $^{241}$Am source. Three dedicated regions were specified for two different track density conditions as follows: (1) Region (i) to accumulate a track density of 300 tracks per 100 $\times$ 100 $\mu$m$^2$, and (2) two dedicated regions (ii) and (iii) to accumulate a track density of 1 $\times$ $10^4$ tracks per 100 $\times$ 100 $\mu$m$^2$.

\begin{table}[H]
\centering
\caption{\label{tab:i} Track density and bleaching condition for dedicated regions on FNTD.}
\smallskip
\begin{tabular}{c c c }
\hline
Regions&Track density per 100 $\times$ 100 $\mu$m$^2$& Beaching condition\\
\hline
Region (i) &300 & Once bleached \\
\hline
Region (ii) & 1 $\times$ $10^4$ & Once bleached \\
\hline
Region (iii) & 1 $\times$ $10^4$ & Twice bleached \\
\hline
\end{tabular}
\label{tab:regions}
\end{table}

Regions (i) and (ii) were used to compare
the results for low and high track densities during the cycles of the irradiation with alpha-particles and optical bleaching. Regions (ii) and (iii) were used to compare different optical bleaching conditions, i.e., region (ii) was bleached under a predetermined bleaching time (the bleaching time is discussed in subsection \ref{exp_sec}) and region (iii) was bleached for twice the initial bleaching time. Hereafter this region will also be referred to as "twice bleached region". The track densities and bleaching conditions are listed in Table \ref{tab:regions}.

\begin{figure}[H]
    \centering
    \includegraphics[width=0.9\linewidth]{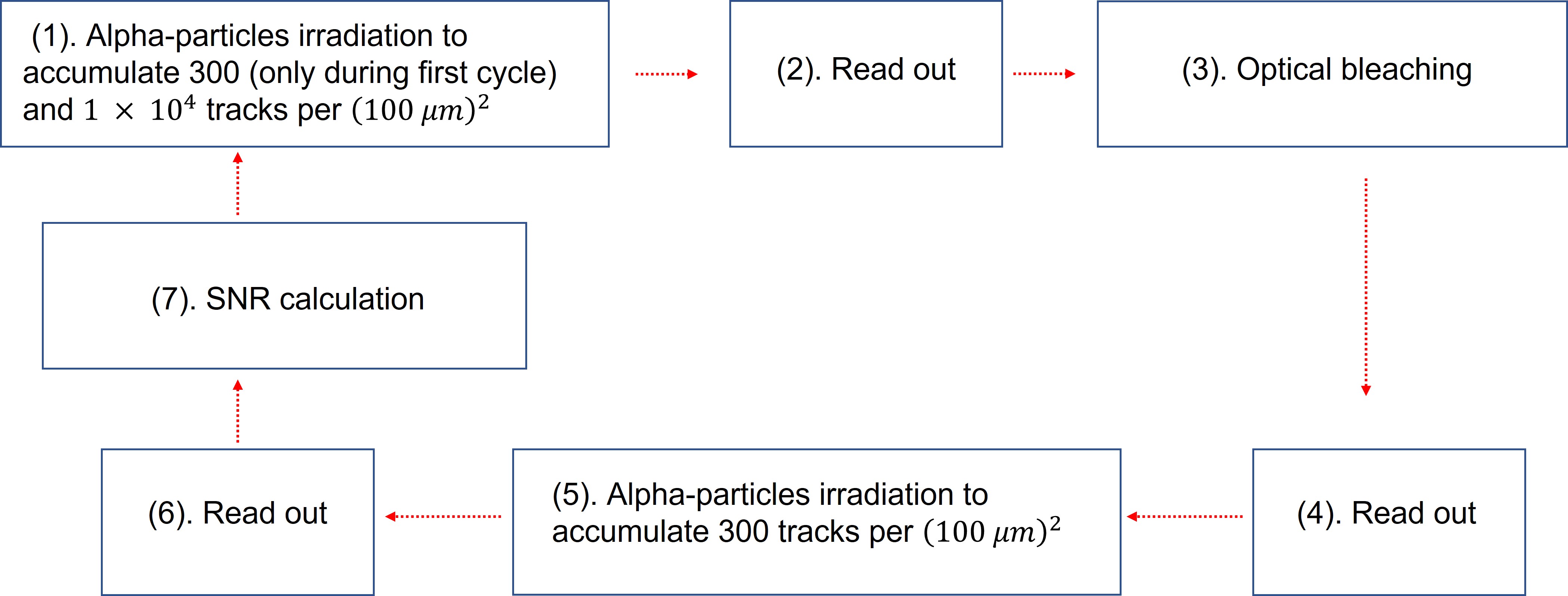}
    \caption{Diagram for illustrating step wise procedure for each cycle of irradiation with alpha-particles and bleaching three dedicated regions on a single FNTD. (1) These regions were irradiated to accumulate a track density of 300 (during the first irradiation-bleaching cycle, afterwards 300 tracks were accumulated during step (5)) and 1 $\times$ $10^4$ tracks per 100 $\times$ 100 $\mu$m$^2$. The track density and bleaching condition for each region are listed in Table \ref{tab:regions}. (2) Tracks of alpha-particles recorded in these regions were imaged using a confocal laser scanning microscope. (3) Optical bleaching of these regions was performed by using a UV laser light. (4) After bleaching, these regions were imaged using a confocal laser scanning microscope. (5) The dedicated regions (i), (ii), and (iii)} were again irradiated to accumulate a track density of 300 tracks per 100 $\times$ 100 $\mu$m$^2$. (6) After irradiation, the dedicated regions were imaged using a confocal laser scanning microscope to obtain the data for SNR calculation. (7) The SNR values were calculated for each region.
    \label{fig:rep_ite}
\end{figure}
The irradiation with alpha-particles and optical bleaching were repeated seven times through a step wise procedure as illustrated in Fig. \ref{fig:rep_ite}. The steps of each cycle are explained as follows. Irradiation with alpha-particles from an alpha source positioned at 2 mm from the surface of the FNTD was used to accumulate a track density of 300 tracks per 100 $\times$ 100 $\mu$m$^2$ and 1 $\times$ $10^4$ tracks per 100 $\times$ 100 $\mu$m$^2$. These dedicated regions were optically bleached using the UV laser system with an output wavelength of 355 nm. A UV laser system used for the optical bleaching is discussed in subsection \ref{exp_sec}. After optical bleaching, these regions were irradiated with alpha-particles to accumulate a track density of 300 tracks per 100 $\times$ 100 $\mu$m$^2$ for the calculation of SNR values. After each cycle of the irradiation with alpha-particles and optical bleaching, the SNR values were calculated as it is a suitable metric to assess the image quality of FNTDs \cite{AKSELROD201835, SYKORA2010631}. The procedure to the calculation of SNR values is discussed in subsection \ref{image_analysis}.

\subsection{Scanning the FNTD}
After each step of irradiation with alpha-particles and optical bleaching, the dedicated regions were imaged using a confocal laser scanning microscope (FV3000RS, Olympus) to obtain the image data sets. A fiber-optic laser with a wavelength of 640 nm and a high numerical aperture (NA), Olympus UPlanFL N 40x/1.30NA oil objective lens, were used to scan the FNTD.  In the beginning of the readout process, the surface of the FNTD was determined manually by changing depth of the focal plane and rotating the focusing knob of the microscope to focus the excitation light from the 640 nm laser. After the determination of the surface, the readout of the FNTD was performed in parallel planes 1 $\mu$m from the surface. The scanning of the focal point was performed under the condition of utilizing 3$\%$ of the maximum laser power. The average laser power during the whole duration of seven irradiation-bleaching cycles was 7.0 $\pm$ 1.1 mW. The raw images obtained with the microscope were sets of sequential fluorescent images of 1024 pixels $\times$ 1024 pixels with dimensions of 106.07 $\times$ 106.07 $\mu$m$^2$ for each field of view.

\subsection{Optical bleaching with laser system} \label{exp_sec}
\begin{figure}[H]
    \centering
    \includegraphics[width=0.9\linewidth]{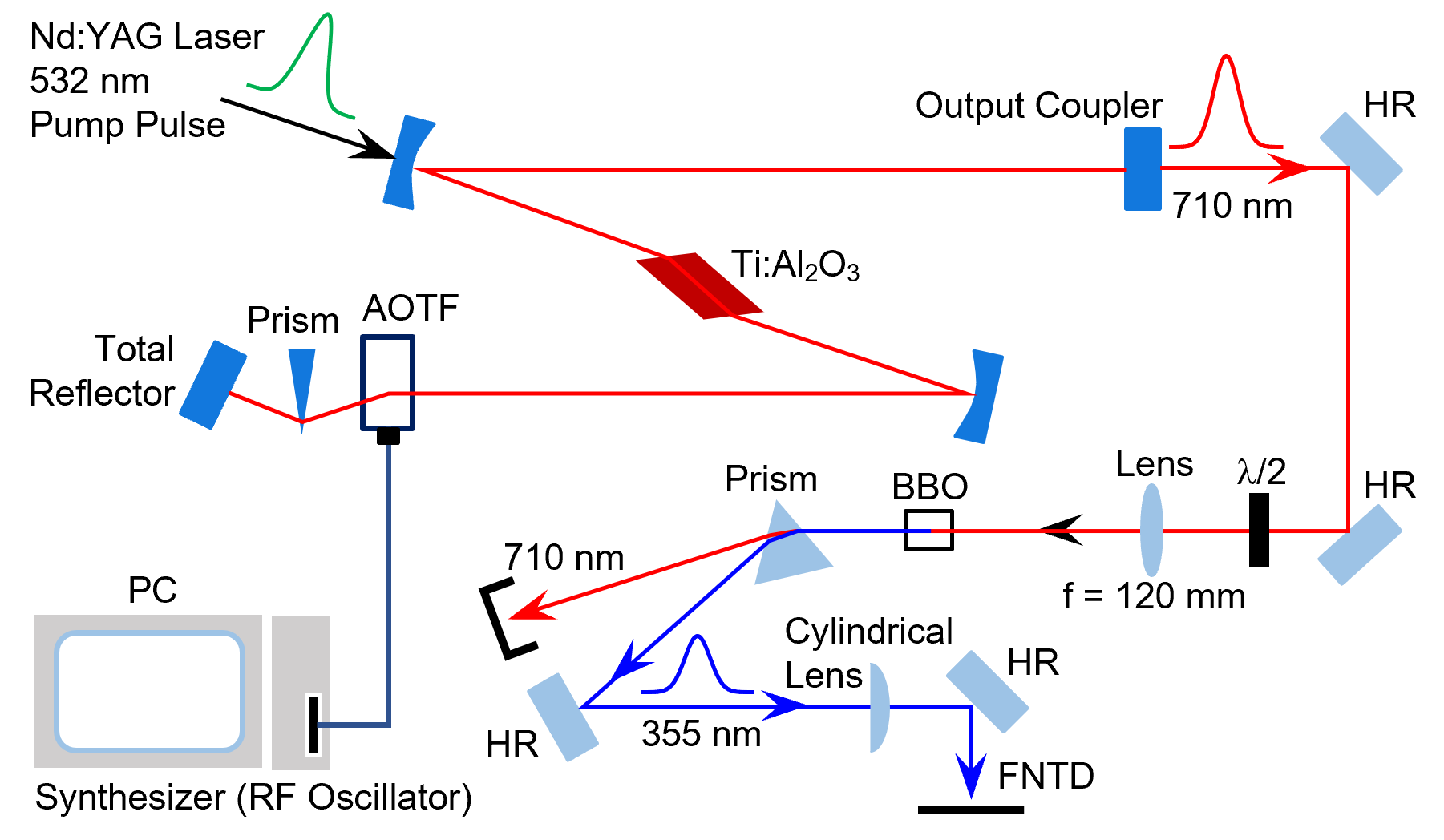}
    \caption{Schematic drawing of the laser system for optical bleaching.
    The Ti:Al${_2}$O${_3}$ laser was pumped with an acousto-optic Q-switched Nd:YAG laser. The output beam was obtained in a second harmonic generation of a pulsed Ti:Al${_2}$O${_3}$ laser in a single nonlinear crystals, $\beta$-BaB${_2}$O${_4}$ (BBO). An AOTF made of TeO${_2}$ was used for the wavelength tuning in the resonator of the Ti:Al${_2}$O${_3}$ laser. The output beam was focused towards the BBO by a lens with a focusing distance of 120 mm. The 355 nm beam emitted from the BBO was separated by a prism from the 710 nm beam. HR is a high reflector, and $\lambda$/2 is half-wave plate.}
    \label{fig:laser_sche}
\end{figure}

For the optical bleaching of FNTDs, a UV laser system with an output wavelength of 355 nm was developed. The output beam was obtained in a second harmonic generation of a pulsed Ti:Al${_2}$O${_3}$ laser in a single nonlinear crystals, $\beta$-BaB${_2}$O${_4}$ (BBO). We used an anti-reflection-coated noncolinear acousto-optic tunable filter (AOTF) made of TeO${_2}$ for the wavelength tuning in the resonator of the Ti:Al${_2}$O${_3}$ laser \cite{Wada96, Saito06}. The Ti:Al${_2}$O${_3}$ laser was pumped with an acousto-optic Q-switched Nd:YAG laser. Details of the resonator construction are given in Ref. \cite{Saito06}. 
The use of AOTF permits arbitrary tuning of the wavelength in the gain region of Ti:Al${_2}$O${_3}$ by computer control. The output energy was 43.3 $\mu$J/pulse at 710 nm, which was measured by a fiber-optic wavemeter (Ocean Optics, Inc.), inputting a radio frequency at 131.2 MHz to the AOTF. The pulsed laser was operated at a repetition rate of 3 kHz. The pulsewidth was approximately 26 ns. The output beam was focused towards the BBO by a lens with a focusing distance of 120 mm. The BBO was angle-tuned to the type I phase-matched angle ($\theta$ = 33.1$^\circ$) for second harmonic generation. The 355 nm beam emitted from the BBO was separated by a prism from the 710 nm beam. Although the 710 nm beam shape was the TEM$_{00}$ (M$^2$ $\approx$ 1.7), the output beam at a wavelength of 355 nm was elliptically distributed. TEM$_{00}$ is the transverse electromagnetic lowest mode Gaussian beam, and M$^2$ is the standard parameter used to determine the quality of experimental beam in compassion to mathematically derived mode of the Gaussian beam. The beam was adjusted in a rectangular pattern (X = 0.466 mm and Y = 0.960 mm) in front of the irradiated sample by use of a cylindrical lens. It was focused in a  spot of 0.45 mm$^2$. The typical energy of the 355 nm beam was 1.7 $\mu$J, 5 mW at 3 kHz. The experimental setup of the laser system used for the optical bleaching of the FNTD is schematically shown in Fig. \ref{fig:laser_sche}.

Since optical bleaching was performed using two-photon absorption photoionization process, the tracks recorded in FNTDs can be erased in a time duration of UV laser light exposure by a factor of (laser power)$^{-2}$ \cite{AKSELROD201835,Akselrod2003}. The recorded tracks in each dedicated region were erased using the 355 nm UV laser light at 5 mW in a time period of at least 8 min. Therefore, optical bleaching can be performed within a time period of 200 $\times$ (laser power)$^{-2}$ min, the laser power is in the unit of mW. As the output power of the laser fluctuates depending on environmental factors such as the temperature, the UV laser irradiation time was tuned to compensate the laser power to bleach the FNTD. For region (iii) under the accumulated track density of 1 $\times$ $10^4$ tracks per 100 $\times$ 100 $\mu$m$^2$, the bleaching time period was twice the estimated bleaching time, i.e, the time period of 2 $\times$ 200 $\times$ (laser power)$^{-2}$ min. 

\subsection{Analysis for the SNR calculation} \label{image_analysis}
The SNR is one of the main criterion to evaluate FNTD performance \cite{AKSELROD201835, SYKORA2010631}. In this study, the fluorescence intensity of tracks and the standard deviation of background fluorescence were calculated as as described below.

A track-detection algorithm, Track selector \cite{Yoshimoto2017}, was employed for the track detection in the acquired sequential fluorescent images. This algorithm can detect the recorded tracks within a certain angular range in the sequential fluorescent images acquired at equal intervals. To detect the tracks within a given angular range, the acquired sequential images of 1024 pixels $\times$ 1024 pixels were converted to binary images of 512 pixels $\times$ 512 pixels. The tracks of alpha-particles within the angular range $|tan\theta{_t}|$ $\leq$ 0.7 were detected, where $\theta{_t}$ is the angle of incidence of alpha-particles to the surface of FNTD. The range of alpha particles with a perpendicular angle of incidence in the FNTD is 15.07 $\mu$m based on calculation by SRIM-2013.00 \cite{ZIEGLER20101818}. However the observed range of alpha particles in the FNTD was
approximately 8 $\mu$m. The possible sources for this attenuation in length are 
refractive indices of glass and FNTD, viscosity of immersion oil, z-pitch
drifting of the confocal laser scanning microscope during the readout process,
the angular distribution of the alpha particles and the air gap between the
source and FNTD. Therefore, we considered 8 $\mu$m depth of the alpha-particles is enough for the measurement. To avoid the multiple detections of a track, clustering was employed to the detected tracks. The segments of tracks within the position space (1.9  $\times$ 1.9 $\mu$m$^2$) and the angular difference within $|tan\theta{_c}|$ $\leq$ 0.3 in the stack of fluorescent images were clustered to reconstruct the trajectory of each detected track, where $\theta{_c}$ is angular difference within a depth of 0.93 $\mu$m (0.93 $\mu$m = 9 pixels) for the clustering of the segments of a track in the acquired sequential fluorescent images.

The mean and standard
deviation of the brightness distribution of the detected tracks and background fluorescence were calculated using the acquired fluorescent image at a depth of 4 $\mu$m from the surface of the FNTD because of the small fluctuation in the dE/d$x$ (stopping power) values. The fluorescence intensity of the detected tracks was calculated as the difference between the mean value of the distributions of the detected tracks and background fluorescence. The fluorescence intensity of the detected tracks and standard deviation of background fluorescence were used to calculate the SNR values after every irradiation-bleaching cycle using a similar definition as in previous works \cite{AKSELROD201835, SYKORA2010631}.

\begin{figure}[H]
    \centering
    \includegraphics[width=0.9\linewidth]{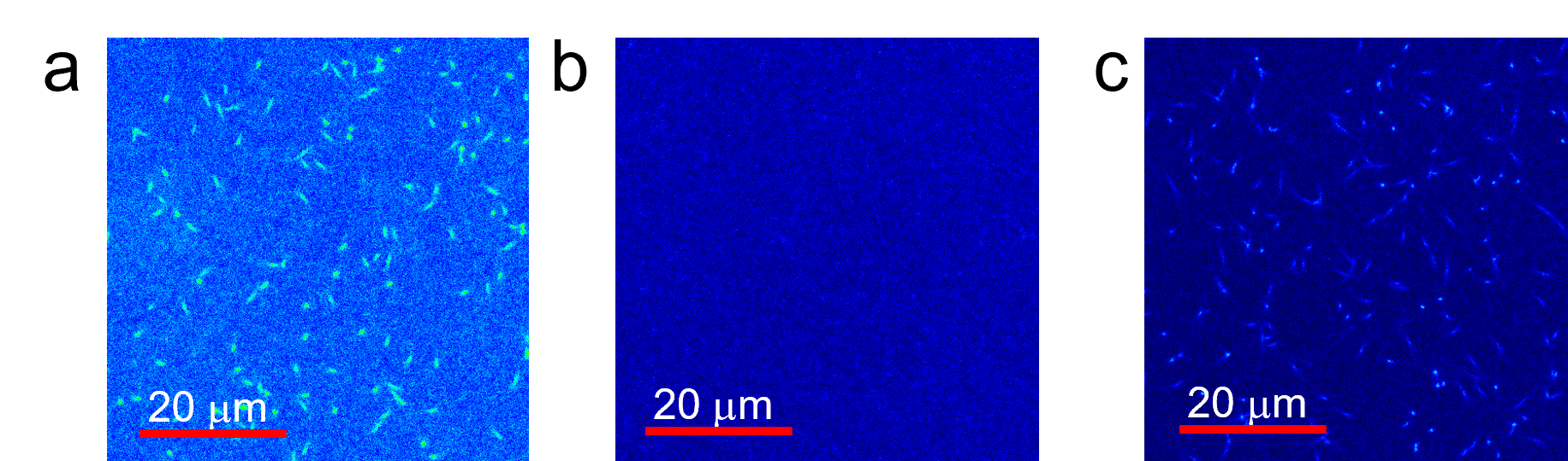}
    \caption{Fluorescent images of ({a}) alpha-particle tracks recorded} in the region (i) under the accumulated track density of 300 tracks per 100 $\times$ 100 $\mu$m$^2$, ({b}) optically bleached region, ({c}) alpha-particle
tracks recorded after the optical bleaching.
    \label{fig:low_tracks}
\end{figure}

\begin{figure}[H]
    \centering
    \includegraphics[width=0.9\linewidth]{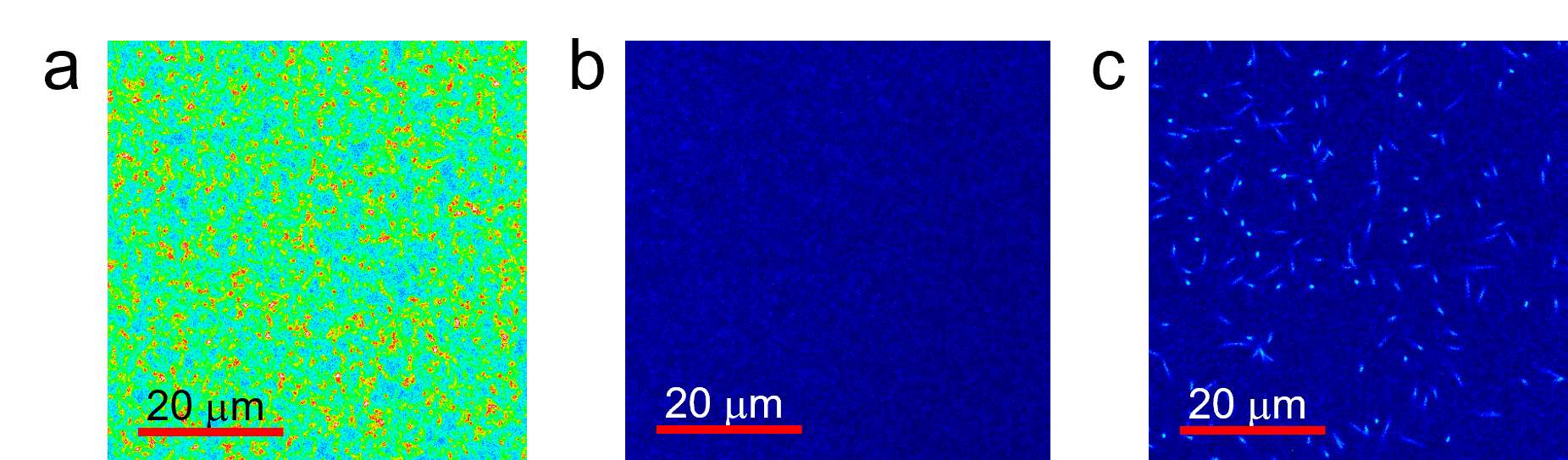}
    \caption{Fluorescent images of ({a}) alpha-particle tracks} recorded in the region (ii) under the accumulated track density of 1 $\times$ $10^4$ tracks per 100 $\times$ 100 $\mu$m$^2$, ({b}) optically bleached region, ({c}) alpha-particle
tracks recorded after the optical bleaching.
    \label{fig:high_tracks}
\end{figure}

\section{Results and Discussion} \label{results}

\begin{figure}[H]
    \centering
    \includegraphics[width=0.9\linewidth]{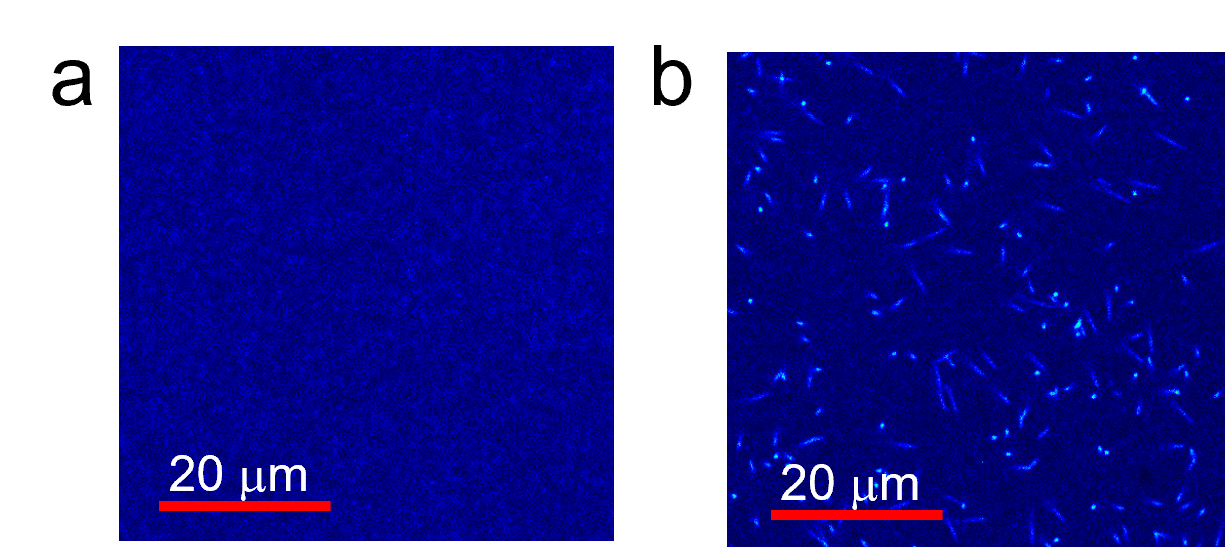}
    \caption{Fluorescent images of ({a}) a region which was twice} bleached under (region (iii)) the accumulated track density of 1 $\times$ $10^4$ tracks per 100 $\times$ 100 $\mu$m$^2$, ({b}) alpha-particle
tracks recorded after the optical bleaching.
    \label{fig:high_tracks_tb}
\end{figure}

Figs. \ref{fig:low_tracks}a and \ref{fig:high_tracks}a show the alpha-particle tracks recorded in the regions (i) and (ii), respectively. The different appearance of the tracks in the regions (i) and (ii) as shown in Figs. \ref{fig:low_tracks}a and \ref{fig:high_tracks}a, respectively, are due to the different track densities. Under a high accumulated track density, the overlapping of tracks induce a large amount of fluorescence. Figs. \ref{fig:low_tracks}b, \ref{fig:high_tracks}b, and  \ref{fig:high_tracks_tb}a show the dedicated regions (i), (ii), and (iii), respectively, which were optically bleached. Figs. \ref{fig:low_tracks}c, \ref{fig:high_tracks}c, and  \ref{fig:high_tracks_tb}b show the alpha-particle tracks were recorded after the optical bleaching.

\begin{figure}[H]
    \centering
    \includegraphics[width=1.0\linewidth]{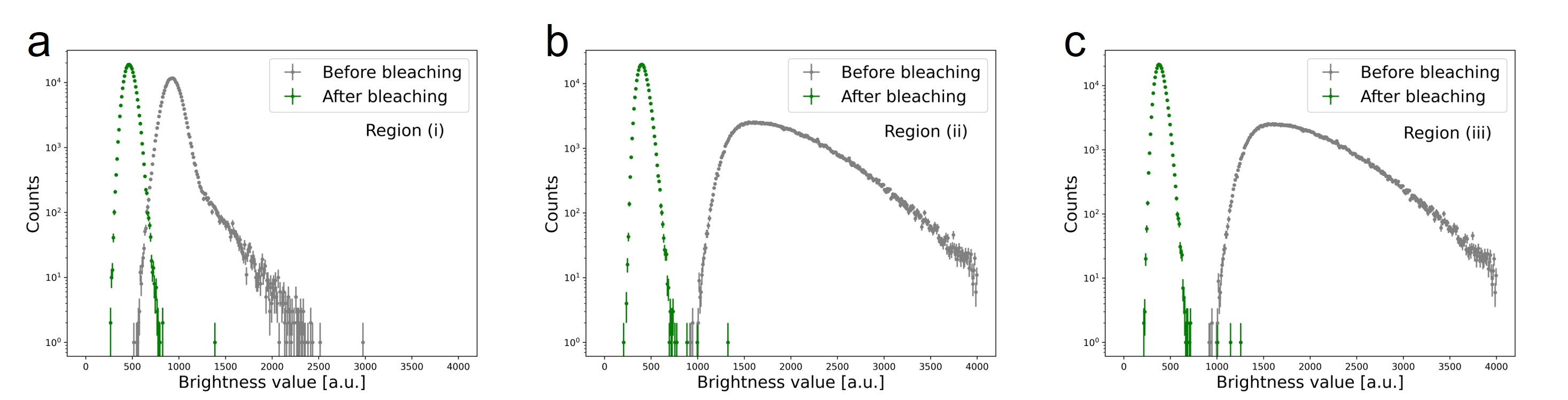}
    \caption{The brightness distributions of acquired fluorescent images before and after the first optical bleaching for (a) region (i), (b) region (ii), and (c) region (iii).  The grey and green marks represent the brightness distribution of all pixels in the acquired fluorescent image before and after the optical bleaching, respectively.}
    \label{fig:bleachin_first_iteration}
\end{figure}

\begin{figure}[H]
    \centering
    \includegraphics[width=1.0\linewidth]{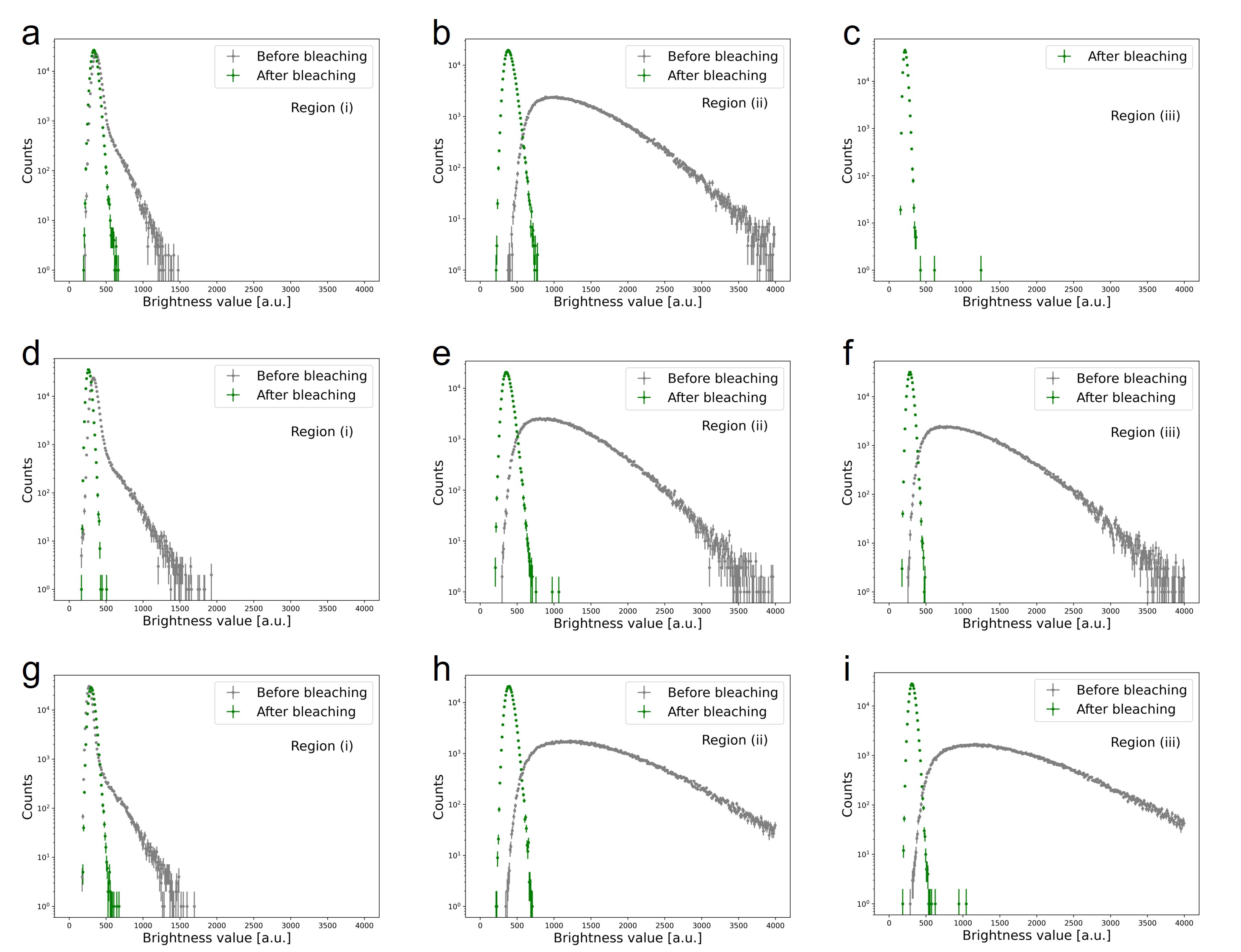}
    \caption{(a) and (b) show the brightness distributions of acquired fluorescent images before and after the second optical bleaching for the regions (i) and (ii), respectively. (d), (e), and (f) Show the brightness distributions of acquired fluorescent images before and after the third optical bleaching for the regions (i), (ii), and (iii), respectively. (g), (h), and (i) Show the brightness distributions of acquired fluorescent images before and after the fourth optical bleaching for the regions (i), (ii), and (iii), respectively. The grey and green marks represent the brightness distribution of all pixels in the acquired fluorescent images before and after the optical bleaching, respectively.}
    \label{fig:bleachin_sectofourth_iteration}
\end{figure}

\begin{figure}[H]
    \centering
    \includegraphics[width=1.0\linewidth]{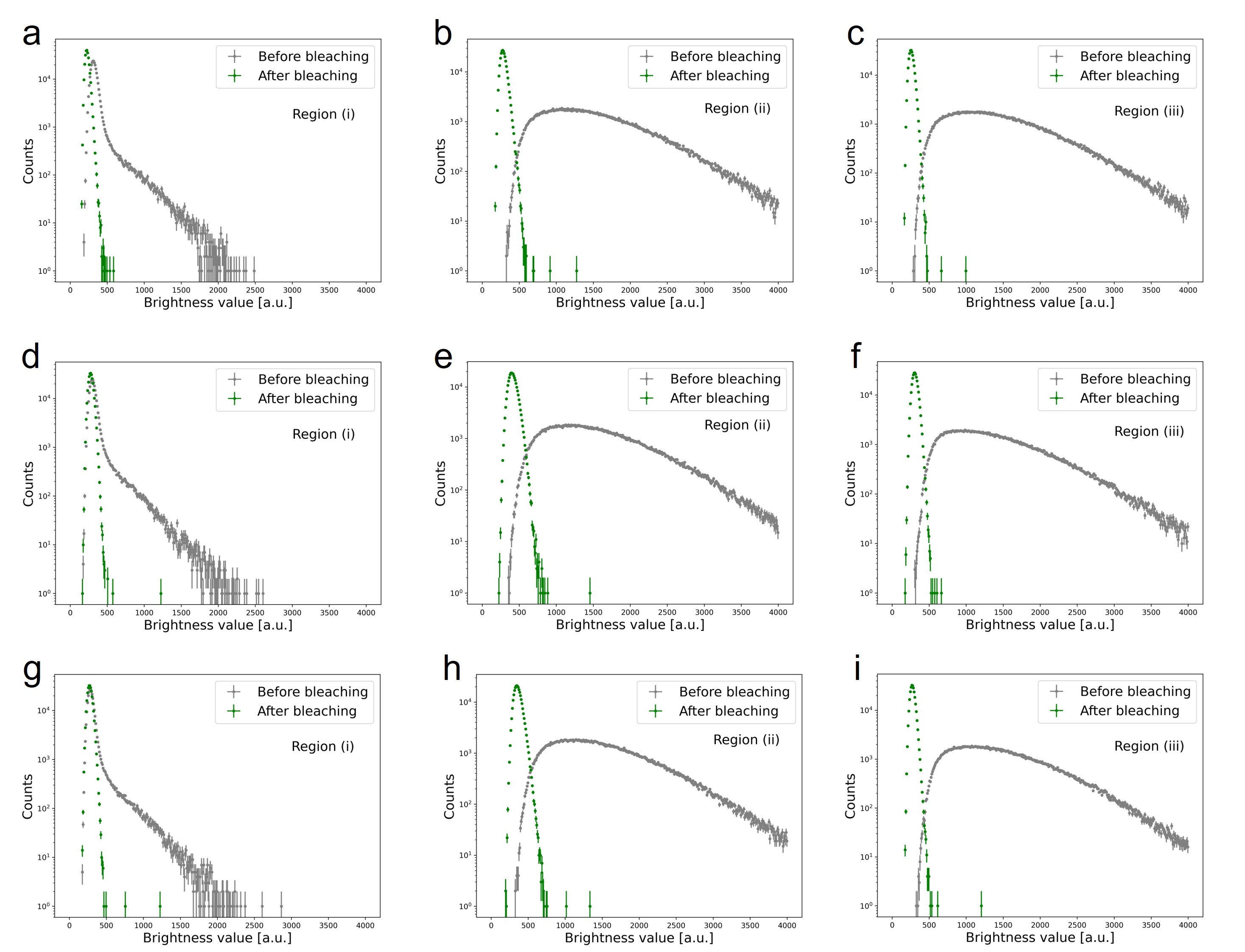}
    \caption{(a), (b), and (c) show the brightness distributions of acquired fluorescent images before and after the fifth optical bleaching for the regions (i), (ii), and (iii), respectively. (d), (e), and (f) Show the brightness distributions of acquired fluorescent images before and after the sixth optical bleaching for the regions (i), (ii), and (iii), respectively. (g), (h), and (i) Show the brightness distributions of acquired fluorescent images before and after the seventh optical bleaching for the regions (i), (ii), and (iii), respectively.  The grey and green marks represent the brightness distribution of all pixels in the acquired fluorescent images before and after the optical bleaching, respectively.}
    \label{fig:bleachin_fifthtoseven_iteration}
\end{figure}

Figs. \ref{fig:bleachin_first_iteration}a, \ref{fig:bleachin_first_iteration}b, and \ref{fig:bleachin_first_iteration}c show the brightness distributions of the fluorescent images after the first bleaching of the regions (i), (ii), and (iii), respectively. Similarly, Figs. \ref{fig:bleachin_sectofourth_iteration}a-c, \ref{fig:bleachin_sectofourth_iteration}d-f, \ref{fig:bleachin_sectofourth_iteration}g-i,
\ref{fig:bleachin_fifthtoseven_iteration}a-c, \ref{fig:bleachin_fifthtoseven_iteration}d-f, and \ref{fig:bleachin_fifthtoseven_iteration}g-i
show the brightness distributions of the fluorescent images after
the second, third, fourth, fifth, sixth and seventh bleaching of regions (i), (ii) and (iii), respectively. The grey and green data points represent the brightness distribution of all pixels in the acquired fluorescent images before and after the optical bleaching, respectively. Due to human error, the brightness distribution before the second
irradiation-bleaching cycle was not measured as shown in Fig. \ref{fig:bleachin_sectofourth_iteration}c. The brightness distributions before bleaching the region (i) have a peak and a tail as shown in Figs. \ref{fig:bleachin_first_iteration}a, \ref{fig:bleachin_sectofourth_iteration}a,  \ref{fig:bleachin_sectofourth_iteration}d,  \ref{fig:bleachin_sectofourth_iteration}g, 
\ref{fig:bleachin_fifthtoseven_iteration}a,  \ref{fig:bleachin_fifthtoseven_iteration}d, and  \ref{fig:bleachin_fifthtoseven_iteration}g. The peak was due the background fluorescence. The tail corresponds to the tracks of alpha-particles with a wider angle of incidence, $|tan\theta{_t}|$ $>$ 0.7. A portion of alpha-particles stop in a depth of 4 $\mu$m depositing more energy because of the high dE/d$x$. Therefore, such tracks appear brighter than the tracks within the angular 
range $|tan\theta{_t}|$ $\leq$ 0.7. The brightness distributions before bleaching the regions (ii) and (iii) during irradiation-bleaching cycles have no background peak which indicate that due to high track density the background is invisible.

Figs. \ref{fig:bleachin_first_iteration}a, \ref{fig:bleachin_sectofourth_iteration}a, \ref{fig:bleachin_sectofourth_iteration}d, \ref{fig:bleachin_sectofourth_iteration}g, \ref{fig:bleachin_fifthtoseven_iteration}a, \ref{fig:bleachin_fifthtoseven_iteration}d, and \ref{fig:bleachin_fifthtoseven_iteration}g, show the brightness distributions before and after bleaching the region (i) during the irradiation-bleaching cycles. The background fluorescence decreased after the first bleaching as shown in Fig. \ref{fig:bleachin_first_iteration}a. From the second until the seventh bleaching cycle, the background fluorescence peaks were overlapping before and after the bleaching which indicate that the luminosity of the FNTD was stable during the irradiation-bleaching cycles. The brightness distributions after the bleaching has no tail and the distributions are almost symmetric, which indicate that the all the recorded tracks in the region (i) were optically bleached. Figs. \ref{fig:bleachin_first_iteration}b, \ref{fig:bleachin_sectofourth_iteration}b, \ref{fig:bleachin_sectofourth_iteration}e, \ref{fig:bleachin_sectofourth_iteration}h, \ref{fig:bleachin_fifthtoseven_iteration}b, \ref{fig:bleachin_fifthtoseven_iteration}e, and \ref{fig:bleachin_fifthtoseven_iteration}h, show the brightness distributions before and after bleaching the region (ii) during the irradiation-bleaching cycles. Similarly, Figs. \ref{fig:bleachin_first_iteration}c,  \ref{fig:bleachin_sectofourth_iteration}f, \ref{fig:bleachin_sectofourth_iteration}i, \ref{fig:bleachin_fifthtoseven_iteration}c, \ref{fig:bleachin_fifthtoseven_iteration}f, and \ref{fig:bleachin_fifthtoseven_iteration}i, show the brightness distributions before and after each bleaching the region (iii) during the the irradiation-bleaching cycles. After first bleaching, the background brightness decreased for the high track density regions (ii) and (iii) and the brightness distributions after the bleaching show that these regions were successfully bleached.

The relatively stable amplitude of the brightness distributions as shown in Figs. \ref{fig:bleachin_first_iteration}, \ref{fig:bleachin_sectofourth_iteration}, and \ref{fig:bleachin_fifthtoseven_iteration} indicate the stability of the luminosity of the FNTD during the irradiation-bleaching cycles. The relatively higher brightness distribution before the first irradiation-bleaching cycle as shown in Fig. \ref{fig:bleachin_first_iteration} may be due to the different background levels of the fresh FNTDs depending on sample.

\begin{figure}[H]
    \centering
    \includegraphics[width=1.0\linewidth]{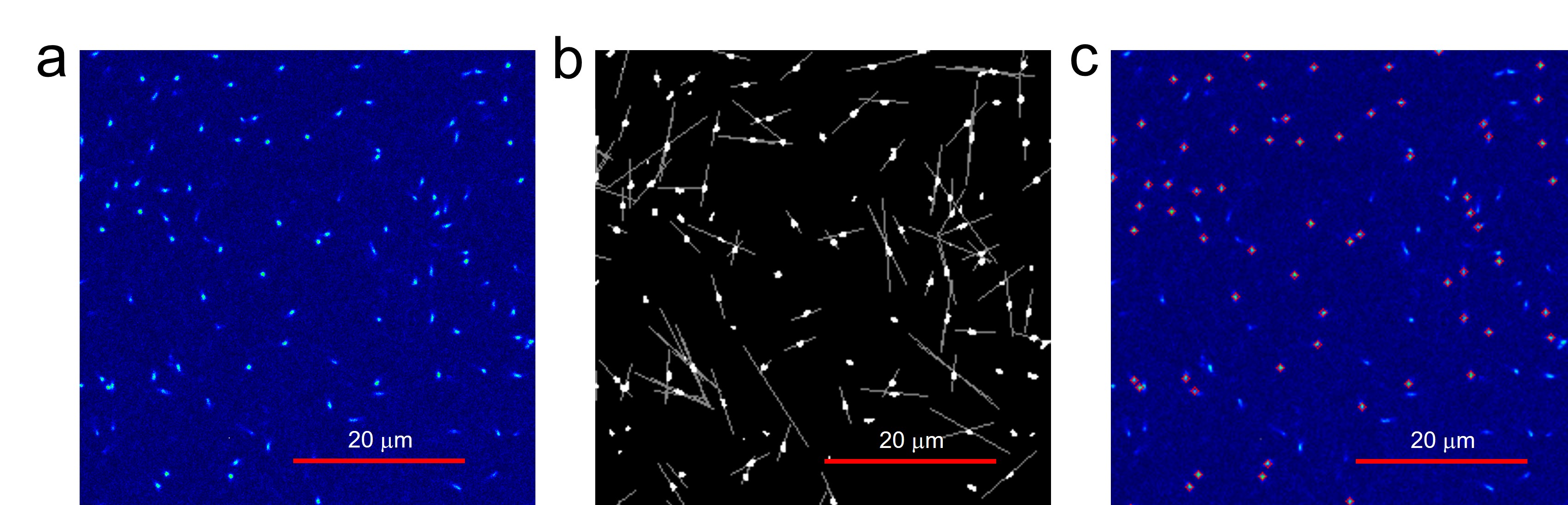}
    \caption{({a}) Shows a fluorescent image of alpha particle tracks recorded in the FNTD, ({b}) shows the corresponding binary image}, the lines indicate the recognized tracks by the track detection algorithm, Track selector, ({c}) shows the detected track-spots in the fluorescent image, where the radius of the each red circle is 2 pixels.
    \label{fig:track_selector}
\end{figure} 

\begin{figure}[H]
    \centering
    \includegraphics[width=0.9\linewidth]{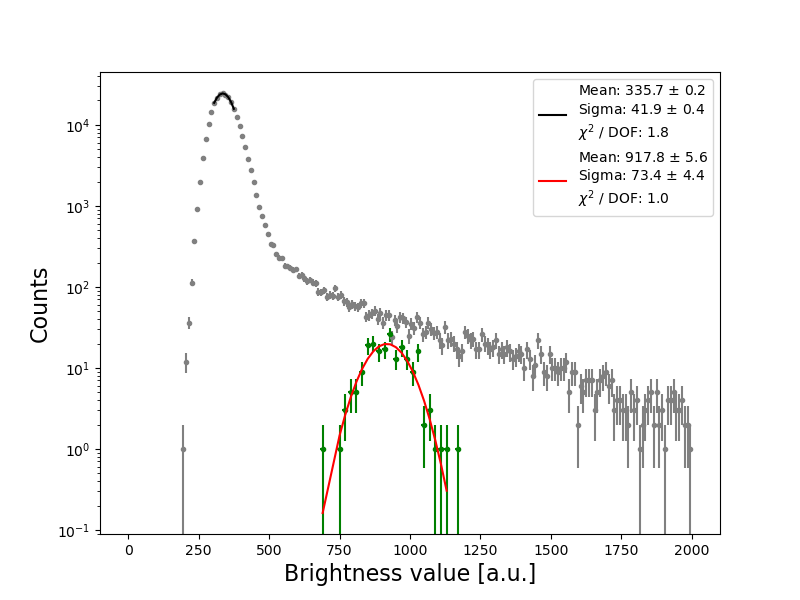}
    \caption{The grey data points represent the brightness distributions of all pixels in the acquired fluorescent image as shown in Fig. \ref{fig:track_selector}a. The green data points show the average brightness
value of the detected track-spots in an area within a circle of radius 2 pixel as shown in Fig. \ref{fig:track_selector}c. The black and red lines show the Gaussian fitting curves for the
background fluorescence and the brightness distribution of detected track-spots, respectively.}
    \label{fig:hist_BG_track_fitting}
\end{figure}

\begin{figure}[H]
    \centering
    \includegraphics[width=0.9\linewidth]{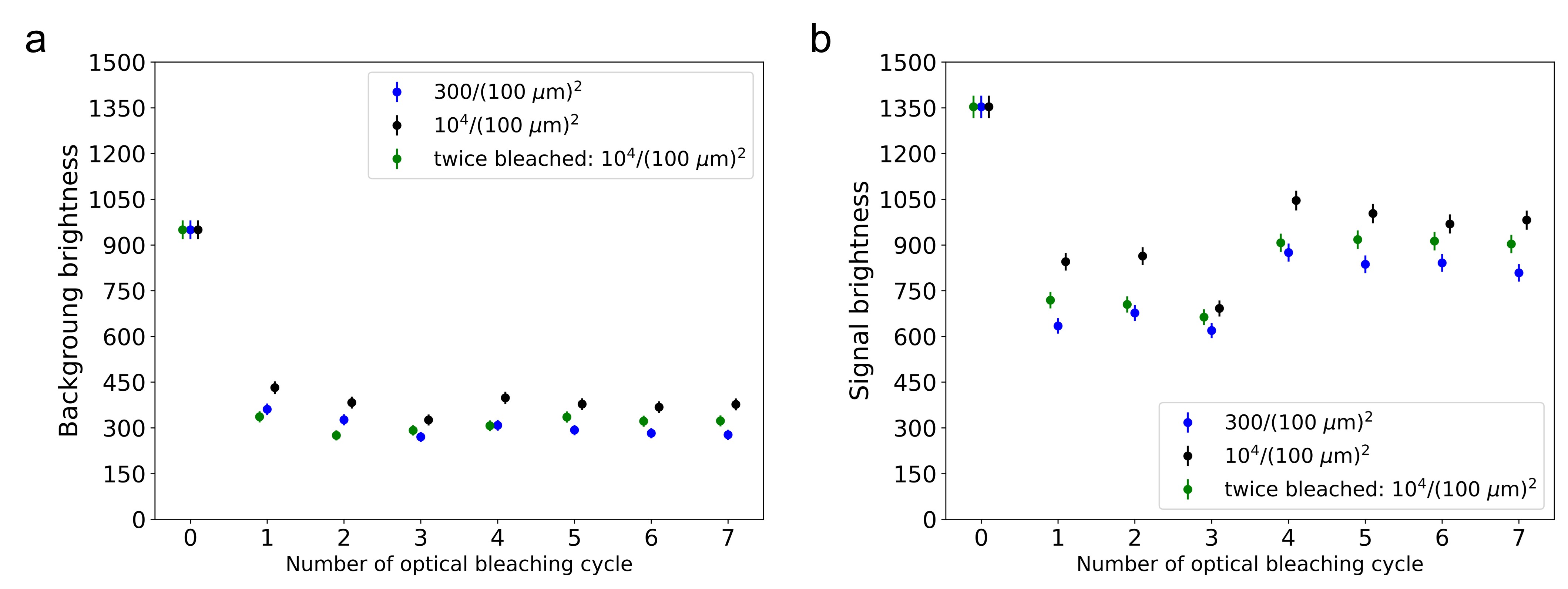}
    \caption{(a) and (b) Show the background and tracks brightness, respectively, during the seven cycles of the optical bleaching.}
    \label{fig:sig_bg_brightness}
\end{figure}

\begin{figure}[H]
    \centering
    \includegraphics[width=0.9\linewidth]{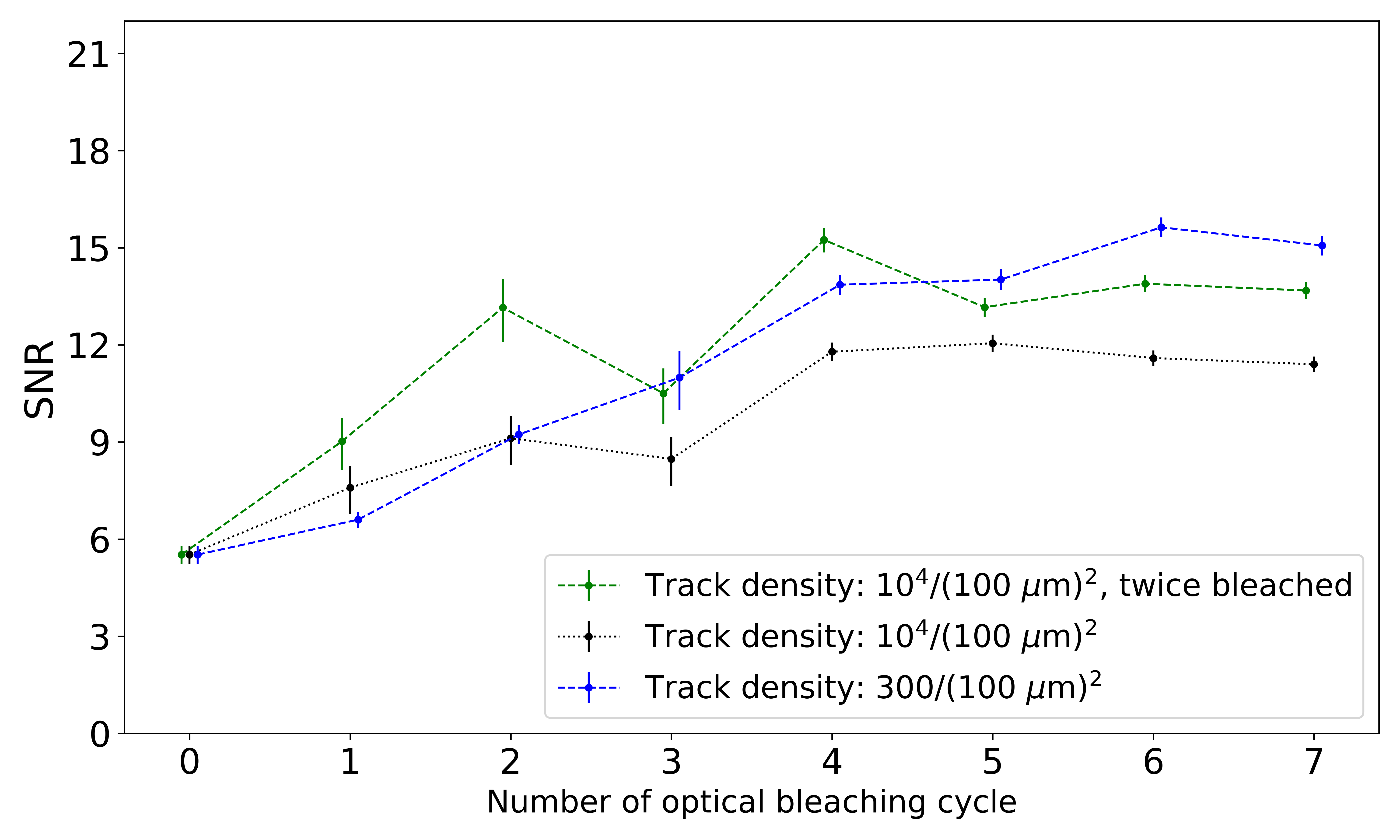}
    \caption{Summary of the SNR calculations} for seven irradiation-bleaching cycles.
    \label{fig:sn_ratio}
\end{figure}

Fig. \ref{fig:track_selector}a shows a single fluorescent image in which the track-spots at that focal plane can be observed. Fig. \ref{fig:track_selector}b is the resized binary image of Fig. \ref{fig:track_selector}a, and the lines indicate the trajectory of detected tracks. For each detected track under the angular range of ($|tan\theta{_t}|$ $\leq$ 0.7) as shown in Fig. \ref{fig:track_selector}c, the average brightness value of the area within a circle of radius 2 pixels around each track was calculated as shown in Fig. \ref{fig:hist_BG_track_fitting}. The mean and standard deviation of the brightness distribution of the tracks and background fluorescence were calculated fitting a Gaussian function, as shown in Fig. \ref{fig:hist_BG_track_fitting}. The fitting range was within the interval of $\pm$ three standard deviations and $\pm$ one standard deviation for the brightness distribution of the tracks and background fluorescence, respectively, to obtain the mean and standard deviation with reasonable reduced $\chi^2$ values for the fitting. The mean and standard deviation of the brightness distributions of detected tracks and background fluorescence after every irradiation-bleaching cycle are shown in Fig. \ref{fig:sig_bg_brightness}. The brightness values
for detected tracks and background decreased compared to those before the irradiation-bleaching
cycles. The SNR values were calculated from the fluorescence intensity of detected tracks (signals) and standard deviation of background fluorescence (noise). 
The results of the SNR calculations are shown in Fig. \ref{fig:sn_ratio}. The initial calculated SNR value was 5.7 $\pm$ 0.2.

As shown in Fig. \ref{fig:sn_ratio}, the SNR for region (i) was higher than the region (ii)).
On the other hand, for the high track density regions (ii) and (iii), the SNR for the twice bleached region (region (iii)) was always higher than the region (ii). Based on these results, we conclude that increasing the bleaching time yields
higher SNR values up to a maximum where there is no subsequent increase within measurement
error. Therefore, to erase the high accumulated track density recorded in FNTDs long UV exposure would be required. A decrease in the SNR values was observed for regions with high track density after the third cycle as shown in Fig. \ref{fig:sn_ratio}. This behaviour was due to the laser power fluctuation of the confocal laser scanning microscope.

The error bars as shown in Fig. \ref{fig:sn_ratio} indicate the combined uncertainty taking into consideration
the statistical and systematic uncertainties. The statistical errors were calculated by the error propagation based on the fitting error of the mean and standard deviation of the distributions. The systematic errors were calculated by using the monthly-based recorded laser power of the confocal microscope. These measurements were performed in the time duration of two months. The relatively large error bars in the SNR for the first three cycles were due to laser power fluctuation of the confocal laser scanning microscope. The bleaching results, shown in Fig. \ref{fig:sn_ratio}, suggest that the tracks recorded in FNTDs can be erased several times using the optical bleaching method, and hence can be re-used at least seven times. 

\section{Conclusion}
In this study, we presented the feasibility of the reusability for FNTDs using seven cycles of irradiation with alpha-particles and optical bleaching. The optical bleaching of the FNTD using a 355 nm UV laser light was performed for the accumulated track densities of 300 and 1 $\times$ $10^4$ tracks per 100 $\times$ 100 $\mu$m$^2$. The SNR values were calculated after each cycle of the irradiation with alpha-particles and optical bleaching of regions with different accumulated track densities and bleaching conditions. Increasing the bleaching time reduces the background effectively and yields better SNR. Based on the results reported in this paper, it is concluded that FNTDs can be re-used at least seven times and can be utilized for the applications where the required track density is approximately of the order $10^4$ tracks per 100 $\times$ 100 $\mu$m$^2$. Particularly the methods here described would be
advantageous for neutron imaging, but also for other applications involving FNTDs.

\section{Declaration of competing interests}
The authors declare that they have no known competing financial interests or personal relationships that could have appeared to influence the work reported in this paper.

\section{Acknowledgments}
The authors would like to thank the RIKEN CBS-Olympus Collaboration Center for the technical assistance with confocal image acquisition. We would like to thank Masahiro Yoshimoto for his technical support related to the Track selector. The irradiation with alpha-particles from the $^{241}$Am source was performed at the Nishina Center, RIKEN. The authors would like to thank Naotaka Naganawa from Nagoya University for fruitful discussions and Yukiko Kurakata of the High Energy Nuclear Physics Laboratory at RIKEN for providing administrative support for the entire project.

 \bibliographystyle{elsarticle-num} 
 \bibliography{cas-refs}





\end{document}